\documentclass[12pt]{iopart}
\usepackage{iopams}
\usepackage{graphicx}
\expandafter\let\csname equation*\endcsname\relax
\expandafter\let\csname endequation*\endcsname\relax
\usepackage{amsmath}

\begin{document}

\title[Thermoelectric power of overdoped Tl2201 crystals]{Thermoelectric power of overdoped Tl2201 crystals: Charge density waves and $T^1$ and $T^2$ resistivities}

\author{J R Cooper$^1$, J C Baglo$^2$, C Putzke$^3$\footnote{Present address: Max-Planck-Institute for the Structure and Dynamics of Matter,
Luruper Chaussee 149, Geb. 99 (CFEL), 22761 Hamburg, Germany}, A Carrington$^4$}

\address{$^1$ Cavendish Laboratory, University of Cambridge, J. J. Thomson Avenue, Cambridge, CB3 0HE, United Kingdom}
\address{$^2$ D\'{e}partement de physique and Institut quantique, Universit\'{e} de Sherbrooke, Sherbrooke, Qu\'{e}bec, Canada}
\address{$^3$ Institute of Materials, \'{E}cole Polytechnique F\'{e}d\'{e}ral de Lausanne, Lausanne, Switzerland}
\address{$^4$ H.\ H.\ Wills Physics Laboratory, University of Bristol, Bristol, BS8 1TL, United Kingdom}
\ead{jrc19@cam.ac.uk, Jordan.Baglo@USherbrooke.ca}

\maketitle

\begin{abstract}
We report measurements of the in-plane thermoelectric power~(TEP) for an overdoped~(OD) crystal of the single layer cuprate superconductor Tl$_2$Ba$_2$CuO$_{6+x}$~(Tl2201) at several hole concentrations~($p$), from 300 or 400~K to below the superconducting transition temperature~($T_c$). For $p$ = 0.192 and 0.220, small upturns in the TEP below 150~K are attributed to the presence of charge density waves ~(CDW) detected by resonant inelastic X-ray scattering studies. This suggests that measurement of the TEP could provide a simple and effective guide to the presence of a CDW. Over a certain temperature range, often strongly restricted by the CDW, the TEP is consistent with the Nordheim-Gorter rule and the $T^1$ and $T^2$ terms in the in-plane resistivity of similar crystals observed below 160~K. Two scenarios in which the $T^1$ scattering term is uniform or non-uniform around the Fermi surface are discussed. As found previously by others, for uniform scattering the $T^1$ terms give scattering rates~($\tau^{-1}$) at lower $p$ that are somewhat larger than the Planckian value $k_BT/\hbar$ and fall to zero for heavily OD crystals. Near 160~K, $\tau^{-1}$ from the $T^2$ terms corresponds to the Planckian value.

\end{abstract}

\vspace{2pc}
 \noindent{\it Cuprates, Thermoelectric power, Charge density waves, Electrical resistivity}

\submitto{\SUST}

\section{Introduction}
Identifying the pairing mechanism in cuprate superconductors and the origin of the characteristic $T$-linear behaviour of the in-plane electrical resistivity, are still controversial and unsolved questions that may turn out to be closely related. In this regard single crystals of Tl$_2$Ba$_2$CuO$_{6+x}$~(Tl2201) are of great interest because their superconducting transition temperature~($T_c$) can be progressively reduced from 90~K to zero by increasing the oxygen content,~$x$, and they have lower in-plane residual resistivity than other cuprates such as La$_{2-x}$Sr$_x$CuO$_4$ (LSCO) and Bi$_2$Sr$_2$CuO$_{6+\delta}$ that can be overdoped~(OD). Furthermore, any van Hove singularity is well away from the Fermi energy~\cite{Putzke2021}. Quantum oscillation (QO) experiments, e.g.~\cite{RourkeQOTl2201} show that heavily OD crystals with $T_c\leqslant$ 26~K have a large Fermi surface. There appears to be no work on strongly underdoped~(UD) Tl2201 crystals, with difficulties caused, at least partly, by the volatility and toxicity of thallium compounds, but this does not affect our conclusions. Recently, charge density waves~(CDW) have been detected below 150~K in a resonant inelastic X-ray scattering~(RIXS) study of OD Tl2201 crystals~\cite{TlCDW} with higher $T_c$. The question as to whether this can account for the unusual evolution of the Hall number as the hole doping~($p$) is reduced~\cite{Putzke2021} is still being debated~\cite{Hussey2023}. In this paper we report measurements of the in-plane thermoelectric power~(TEP), $S_a(T)$ for a single crystal from the same preparation batch~(Oct. 2014) as in the RIXS and Hall effect work. We argue that the CDW gives a small but distinct upturn in $S_a(T)$ below 150~K. This conclusion is reinforced by the observation of similar upturns in all polycrystalline samples~\cite{OCT1992} with $T_c\geq29$~K. Such data usually agree reasonably well with measurements of $S_a(T)$ on single crystals.

We discuss in-plane electrical resistivity $\rho_a(T)$ data~\cite{Putzke2021} for Tl2201 crystals and show that the TEP is consistent with the $T^1$ and $T^2$ terms in $\rho_a(T)$ observed below 160~K and the Nordheim-Gorter~(NG) rule. Corresponding scattering rates~$\tau^{-1}$ are compared with the Planckian rate~$k_BT/\hbar$~\cite{Zaanen2004,Legros2019,Hartnoll2022}. We estimate $\tau^{-1}$ at low $T$ from $\rho_a(0)$ and raise the question as to whether the residual low-$T$ specific heat of OD Tl2201~\cite{Wade1994} is caused by standard pair-breaking in a $d$-wave superconductor~\cite{LeeHone2020,Broun2013} combined with a decrease in the strength of the pairing interaction~\cite{Storey2007,Scalapino2017}, or whether it is caused by the pairing interaction $V_{kk'}$ being sufficiently anisotropic so that some electrons are not paired. Namely if the $T^1$ scattering term is much weaker on some parts of the Fermi surface does this imply that $V_{kk'}$ is much smaller there?

\section{Results and analysis}

\subsection{Effect of CDW on the TEP}
Above $1.1~T_c$ the TEP of OD cuprates generally varies approximately as $a + b(T/300)$, where $a \simeq 1$ to $2~\mu$V/K and $b$ changes from $\simeq -1$ to $-10~\mu$V/K as $p$ is increased from 0.16, the value for maximum $T_c$, to the upper edge of the superconducting dome~\cite{OCT1992} while for UD samples it is usually positive and much larger~\cite{OCT1992}. TEP data were taken in an earlier, independent research project before the RIXS studies, using the method described in the appendix. One crystal was used for the TEP work and re-annealed to give successive $T_c$ values of 26.5~K, 29~K, 67~K and 88~K, annealing conditions are given as Supplementary Information~\cite{SuppMat}. A small correction $\simeq +0.1\mu$V/K below 50~K has been made for the $T_c$ = 29~K and 26.5~K data using the phosphor bronze results shown in the appendix. AC susceptibility data for various crystals of Tl2201 with small AC fields parallel and perpendicular to the CuO$_2$ layers after different annealing treatments are reported as Supplementary Information~\cite{SuppMat}. They reveal an unexpectedly complex behaviour and give evidence for two oxygen diffusion processes with very different time scales. It is probable that one or both of these are influenced by the concentration of Cu ions on Tl sites. This data could be useful for future studies. TEP data for polycrystalline samples~\cite{OCT1992} with $T_c$ = 84~K, 82~K, 59~K, 29~K and less than 4~K, provide important confirmation of the present work.

\begin{figure}
  \includegraphics[width=80mm,height=80mm]{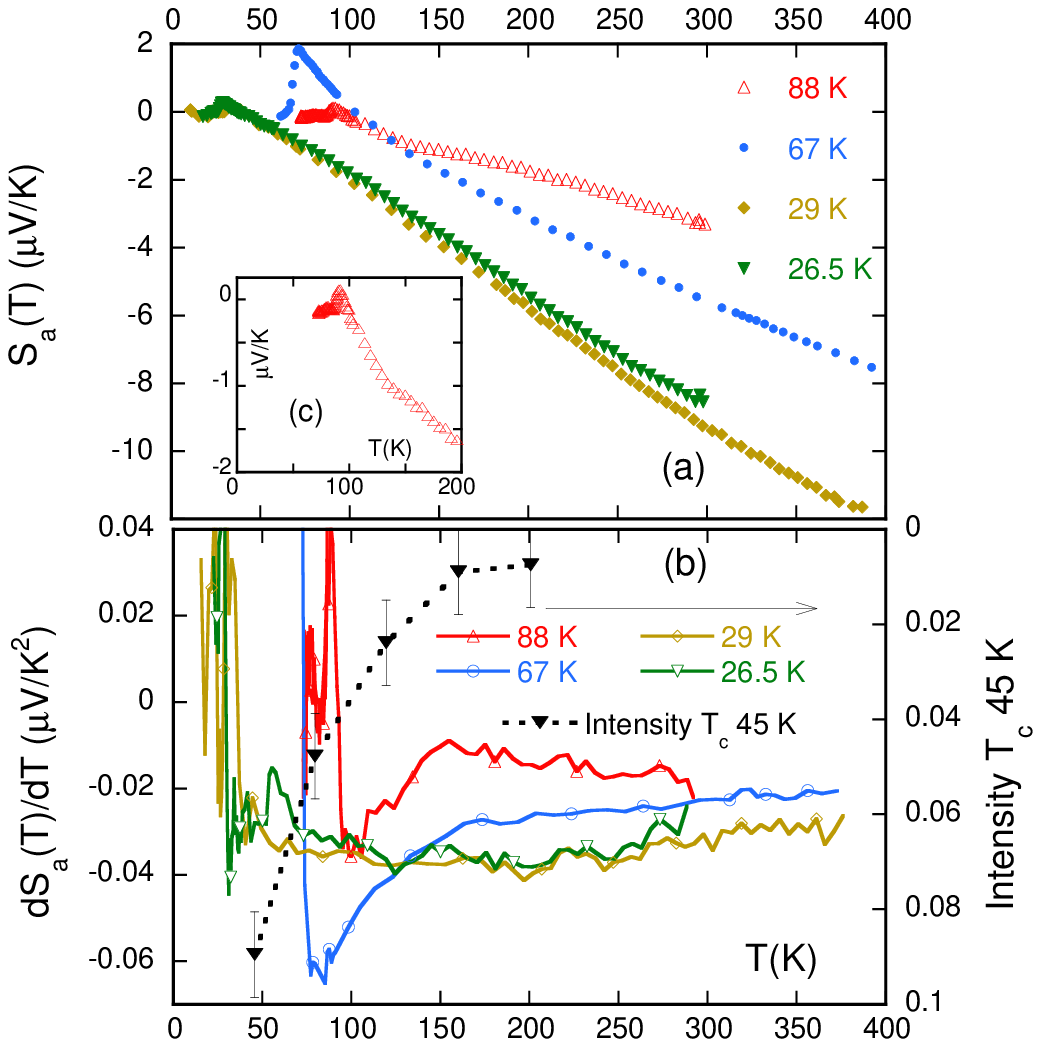}
  \caption{(a) In-plane thermoelectric power $S_a(T)$ for the same Tl2201 crystal with 4 different $T_c$ values. $p$ =0.192, 0.220, 0.271 and 0.274 for $T_c$ = 88~K, 67~K, 29~K and 26.5~K respectively determined from the same modified formula used in Ref.~\cite{Putzke2021} that accounts for the Fermi surface area obtained from quantum oscillation studies~\cite{RourkeQOTl2201} for $T_c\leq26$~K. (b) Left-hand scale, corresponding derivative plots $dS_a/dT$, and right-hand scale RIXS intensity \textit{vs} $T$ for a crystal with $T_c$ =~45~K from~\cite{TlCDW}. The CDW causes a downturn in $dS_a/dT$ below 150~K, upturns at lower $T$ are caused by the onset of superconductivity. Before forming the derivatives, $S_a(T)$ and $T$ data were smoothed over 5 points, typically $\pm10$~K, 10 or 20~K above $T_c$ and $\pm4$~K nearer $T_c$. This reduced the scatter without causing other changes. (c) Inset, expanded view for $T_c$ = 88~K showing a clear anomaly at 145~K.}
 \label{TEPall}
\end{figure}

Fig.~\ref{TEPall}(a) shows that the in-plane TEP $S_a(T)$ for the same Tl2201 crystal with $T_c$ values of 88~K and 67~K has a small upturn below $\sim$150~K, where there is also a clear change in the derivative $dS_a/dT$, as shown in Fig.~\ref{TEPall}(b), but there is no such effect for $T_c=26.5$~K. For $T_c=29$~K there is marginal evidence for a CDW with an onset near 40~K as indicated in Fig.~\ref{SoverTandNG}. All this correlates well with RIXS measurements on Tl2201 crystals from the same preparation batch for which CDWs were detected for $T_c$ values of 56~K and 45~K but not for 22~K~\cite{TlCDW}. TEP data for $T_c$ = 88~K, on an expanded scale in Fig.\ref{TEPall}(c), show a particularly clear anomaly near 145~K.
 \begin{figure}
  \includegraphics[width=80mm,height=100mm]{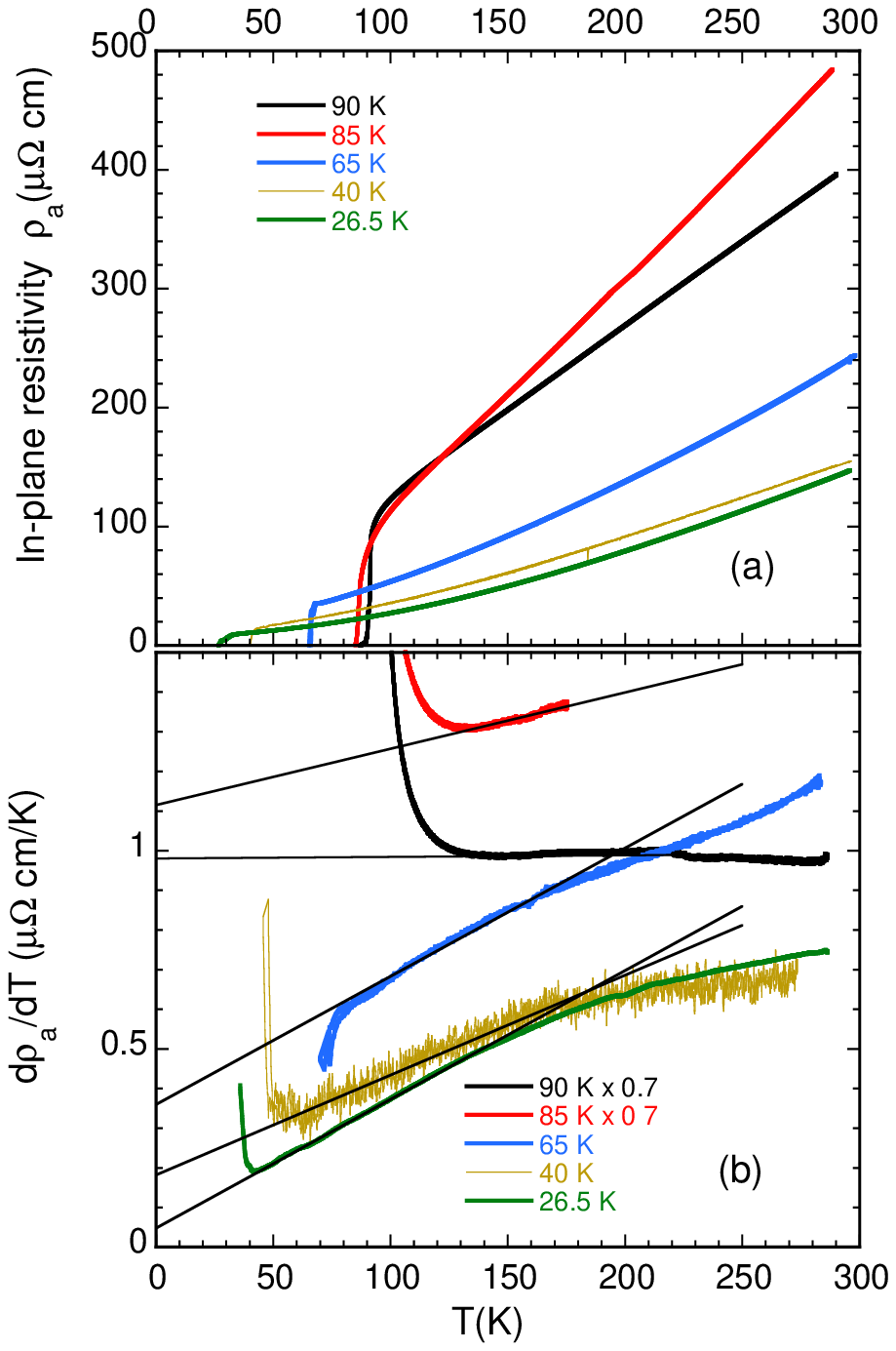}
  \caption{(a) In-plane resistivity data $\rho_a(T)$ for Tl2201 crystals from the same batch, taken from~\cite{Putzke2021}. The crystal with $T_c$ = 85~K is probably underdoped~\cite{Putzke2021} and not considered further here. The $p$ values for the other crystals are 0.183, 0.227, 0.256 and 0.274 for $T_c$ = 90~K, 65~K, 40~K and 26.5~K respectively. (b) Corresponding derivative plots and straight lines expected when $\rho_a(T) = \alpha_0 +\alpha_1T+\alpha_2T^2$. Data above 175~K for $T_c$ = 85~K were too noisy to give accurate derivatives. The straight lines show the results of least squares fits.}
 \label{resdat}
\end{figure}
The CDW onset temperature for $T_c$ = 45~K is 150~K as shown by the RIXS data on the right hand scale of Fig.~\ref{TEPall}(b) but unfortunately the onset temperature of 160~K shown in  Ref.~\cite{TlCDW} for $T_c$ = 56~K is only an upper bound. Although better overlap of $T_c$ values would have been desirable it is very probable that the changes in $S_a(T)$ and $dS_a/dT$ near 150~K shown in Fig.~\ref{TEPall}(a),(b) and (c) as well as at various temperatures for the polycrystalline data in Ref.~\cite{OCT1992} are caused by a CDW. This agreement is potentially useful because it shows that the TEP is a sensitive probe of the CDW and could be used to study its pressure dependence and possibly its magnetic-field dependence. Data for the OD88 crystal and the OD84 polycrystalline sample~\cite{OCT1992} show that the CDW is still present at lower $p$, which is a new finding.

\subsection{$T^1$ and $T^2$ terms in the resistivity} Resistivity data~\cite{Putzke2021} for five different crystals from the same preparation batch are shown in Fig.~\ref{resdat}(a). As noted in~\cite{TlCDW} there appears to be no effect of the CDW on the in-plane resistivity $\rho_a(T)$. This is unusual since other CDW materials do show resistivity anomalies, but it is reinforced by the derivative plots in Fig.~\ref{resdat}(b). Although there is a change in slope of $d\rho_a(T)/dT$ near 160~K, we do not believe that this is caused by a CDW. It is also present for the crystal with $T_c=$~26.5~K and a crystal from another batch with $T_c=$~17~K as well as for older data scanned in from Ref.~\cite{Tyler1997}.

Such derivative plots~\cite{Hussey2013,Culo2021} are a convenient way of applying the formula
\begin{equation}
\label{rhopoly}
{\rho_a(T) = \alpha_0 +\alpha_1T+\alpha_2T^2}
 \end{equation}
because the range of fit is easily seen by inspection. They suggest that there are two distinct values for the $T^1$ and $T^2$ terms, below and above 160~K where both the slopes and the intercepts in Fig.~\ref{resdat}(b) change. For the crystals with $T_c$ = 65~K, 40~K and 26.5~K, $\alpha_1$ and $\alpha_2$ were found from the intercepts at $T=0$ and the slopes of the straight lines in Fig.~\ref{resdat}(b), while the residual resistivity $\alpha_0$ was then found by fitting the data in Fig.~\ref{resdat}(a) to a second order polynomial in which $\alpha_1$ and $\alpha_2$ were fixed. The coefficients $\alpha$ are given in Table~\ref{rhotab}. (A small correction for a measurement offset of $\sim$1~$\mu\Omega$-cm below $T_c$ has been included where appropriate). Here we focus on the values below 160~K, expecting them to be more closely related to superconductivity. The coefficient $\alpha_1$ is presently of great interest because of the suggestion that the Planckian scattering rate~$\hbar/\tau \approx k_BT$~\cite{Zaanen2004} is the maximum allowed for any material~\cite{Hartnoll2022}, and that it is an important intrinsic property of superconducting cuprates~\cite{Legros2019} and many other correlated metals~\cite{Hartnoll2022}. The Drude formula for the in-plane conductivity of an isotropic cylindrical Fermi surface, $\sigma=ne^2\tau/m^*$, which is a reasonable first approximation for Tl2201, shows that $\alpha_1=$ 1 $\mu\Omega$-cm/K corresponds to $\hbar/\tau \approx 3.3 k_BT$, if the carrier density $n$ = 1.3 holes per unit cell and $m^*=5.2m_e$. Although this value of $m^*$ has been measured by QO only for OD Tl2201~\cite{RourkeQOTl2201}, specific heat data~\cite{Wade1994} show that it only varies weakly with $p$. So $\alpha_1=$ 0.3 $\mu\Omega$-cm/K in Fig.~\ref{resdat}(b) corresponds to $\hbar/\tau= k_BT$. For crystals with $T_c<$~50~K, $\hbar/\tau$ falls below this value because, as found previously~\cite{Hussey2013,Culo2021}, the size of the $T^1$ term falls linearly to zero as $p$ approaches the edge of the superconducting dome. This could be an indication that the anomalous scattering becomes confined to certain parts of the Fermi surface, in contrast to the results for some other cuprates~\cite{Legros2019}. The value of $\alpha_0$ for $T_c$= 26.5~K in Table~\ref{rhotab}, gives $\hbar/(k_B\tau)=19$~K and for a Fermi velocity $v_F \equiv\hbar k_F/m^*=1.7 \times 10^7$ cm/s, a mean free path of 68~nm. Specific heat data for Tl2201~\cite{Wade1994} show a large residual linear term which increases as $p$ approaches the edge of the superconducting dome, as expected when $\hbar/(k_B\tau)$ becomes comparable with $T_c$.

Previously~\cite{Hussey2013} the deviations from a straight line above 160~K for the three most overdoped crystals in Fig~\ref{resdat}(b). were interpreted in terms of a parallel conducting channel and the maximum resistivity of 1800~$\mu\Omega$-cm obtained when the carrier mean free path is equal to the in-plane lattice spacing $a$. However they are still clearly visible for the two most overdoped crystals in Fig.~\ref{resdat}(a) that have resistivities as low as 60~$\mu\Omega$-cm at 160~K and as mentioned already for another crystal from a different preparation batch with $T_c=17$~K. Therefore we have no clear interpretation of this effect. It is worth noting that $\alpha_2$ is rather insensitive to $p$ as also shown in~\cite{Hussey2013}. At 160~K it corresponds to a $T^2$ term in the resistivity of 40~$\mu\Omega$-cm and, making the same assumptions as before, a scattering rate $\hbar/\tau = k_BT$, the Planck condition~\cite{Zaanen2004}. This observation could have ramifications for interpretations of the $T^1$ term arising from the Planck condition. Perhaps $\hbar/\tau \geq k_BT$ might cause the deviations from linearity in Fig.~\ref{resdat}(b) above 160~K. Alternatively the deviations might be connected with the effect of lattice vibrations on electron-electron Umklapp scattering. Electron-electron scattering is rarely studied up to such high temperatures. It is known that the Debye-Waller factor, which reduces the intensity of X-ray diffraction lines, causes the energy gap of semiconductors such as Si to fall as $T$ is increased. It is possible that Umklapp scattering between Brillouin zones is reduced by a similar $T$-dependent factor that becomes more significant as the amplitude of the lattice vibrations increases.

Data for all hole-doped cuprates clearly show that the curvature of in-plane $\rho(T)$ plots increases on the OD side. It is widely accepted that there is a crossover line on the $T-p$ phase diagram~\cite{Hussey2008}, which although ill-defined, increases linearly with $p$ from zero for $p\geq0.19$. It can be interpreted as a coherence temperature~$T_{\textrm{coh}}(p)$ for the charged carriers and when $T\lesssim T_{\textrm{coh}}$ the $T^2$ term dominates. The results for $T\leq$ 160~K in Fig.~\ref{resdat}(b) and Table~\ref{rhotab} are \emph{qualitatively} consistent this is because $\alpha_1$ falls with $p$ while $\alpha_2$ stays constant. For $T\geq$ 160~K this requires the slope of $d\rho_a/dT$ to become constant which is the case above 200~K  for $T_c=$ 30~K~\cite{Hussey2013} and for data in Ref.~\cite{Tyler1997} above 250~K for several $T_c$ values between 0 and 80~K. It is puzzling that this does not hold for $T_c$ = 65~K in Fig.\ref{resdat}(b) suggesting that more data are needed to clarify this point. We note that there can be complications from Cu ions on the Tl sites affecting $p$~\cite{SuppMat} as well as time dependent changes in $T_c$ when a Tl2201 crystal is held at room temperature~\cite{Mackenzie1996} that were also seen in our TEP data for $T_c$ = 26.5~K.

\subsection{Pair breaking \textit{vs} ``not-pairing''}
There seems to be two possible ways of explaining the variation of $\alpha_1$ and $T_c$ with $p$, which we will refer to as scenarios A and B. The more standard one, scenario A, based on the usual $d$-wave form for the pairing interaction, $V_{kk'}$~$\propto$~$(\cos k_x - \cos k_y)( \cos k'_x - \cos k'_y)$, has been thoroughly analysed and discussed in Refs.~\cite{LeeHone2020,Broun2013} and related papers where the importance of small angle scattering was emphasised. Here $T_c$ falls as $p$ is increased because the pairing interaction falls~\cite{Storey2007,Scalapino2017} and hence $(\hbar/\tau)/\Delta$ where $\Delta$ is the superconducting energy gap, is larger. The theory is very similar to that used to understand the effect of magnetic impurities in classical superconductors. Experimental evidence for the relation between $T_c$ and $\rho_{res}$ is summarised in Ref.~\cite{Rullieruni2000} for both UD and OD cuprates. A second, less conventional one, scenario B, involves a different variation of the pairing interaction around the Fermi surface and is considered later. Scenario A could be referred to as ``pair-breaking'' and B as ``not-pairing''. For our data set, the values of $\alpha_0$ for the OD65, OD40 and OD26 crystals in Table~\ref{rhotab} are very close to that in Ref.~\cite{Broun2013} for their OD25 crystal measured using microwaves.

\subsection{Nordheim-Gorter analyis}
In standard transport theory, when there are two or more independent scattering mechanisms, for example electron-impurity and electron-phonon scattering, giving contributions $\rho_0$ and $\rho_1$ to the resistivity, the total resistivity $\rho=\rho_0 + \rho_1$, which is known as Matthiessen's rule~(MR). For ordinary metallic alloys MR is a good first approximation although there are systematic deviations from MR that are well-documented~\cite{Cimberle_Rizzuto1974}. There is considerable experimental evidence for Zn-doped YBCO and LSCO that MR applies to the $T$ - independent impurity scattering and the $T^1$ scattering terms in the cuprates~\cite{Cooper1991,Walker1995,Uchida1996}. The equivalent way of calculating the TEP is via the Nordheim-Gorter~(NG)~\cite{MacD1962} rule:
\begin{equation}
\label{NGrule}
 { S_{NG} =( \rho_0S_0+\rho_1S_1+\rho_2S_2 + \cdots) /(\rho_0 +\rho_1 +\rho_2 + \cdots)}
 \end{equation}
where $S_0$ and $S_1$ etc. are the corresponding contributions to the TEP. If the electron diffusion TEP is dominant, and there are no small energy scales, these will be proportional to $T$, but with different signs and magnitudes governed by the energy dependence of the particular scattering mechanism. As shown in Table~\ref{rhotab} the $T_c$ values in Fig.~\ref{TEPall} are close enough to some of those in Fig.~\ref{resdat}(a) so the coefficients $\alpha$ can be used directly to calculate the TEP via Eq.~\ref{NGrule}.
\begin{table}
\caption{Values of $\alpha$ given by the straight lines below 160~K in Fig.~\ref{resdat}(b) plus a constrained fit to the data in Fig.~\ref{resdat}(a) (see text), i.e. $\rho(T) = \alpha_0 + \alpha_1T +\alpha_2T^2$. The carrier mean free path~(mfp) is obtained from $\alpha_0$. $T_c^{\textrm{TEP}}$ values are from Fig.~\ref{TEPall}(a). $S_0$, $S_1$ and $S_2$ are from fits to Eq.~\ref{NGrule}, but cannot be determined independently (see text). The errors in $\alpha$ values are $\pm10\%$ from geometrical uncertainties. }
\label{rhotab}
 \begin{tabular}{@{}lllllllll}
\br
$T_c$ &$\alpha_0$ & $\alpha_1$ &  $\alpha_2$ x 1000 & mfp&$T_c^{\textrm{TEP}}$&S$_0/T$&S$_1/T$&S$_2/T$\\
(K)   &  ($\mu \Omega$-cm) & ($\mu \Omega$-cm/K) & ($\mu \Omega$-cm)/K$^2$)&(nm)& (K)&($\mu$V/K$^2$)& ($\mu$V/K$^2$)&($\mu$V/K$^2$)\\
 \mr
 90       &$(6.4\pm0.4)$           &1.37                           &0.125& --  & 88 & 0.132& $-0.0115\pm6\%$ & --\\
 65       &6.3                     &0.313                          &1.82 & 65  & 67 &0.128& --     & $ -0.030\pm10\%$\\
 40       &6.9                     &0.17                           &1.34 & 60  & -- & -- & --    &  --\\
 26.5     &6.0                    &0.053                          &1.6  & 68 & 26.5 & 0.016& --     & $- 0.033\pm3\%$\\
 \br
\end{tabular}
\end{table}

\begin{figure}
\includegraphics[width=80mm,height=70mm]{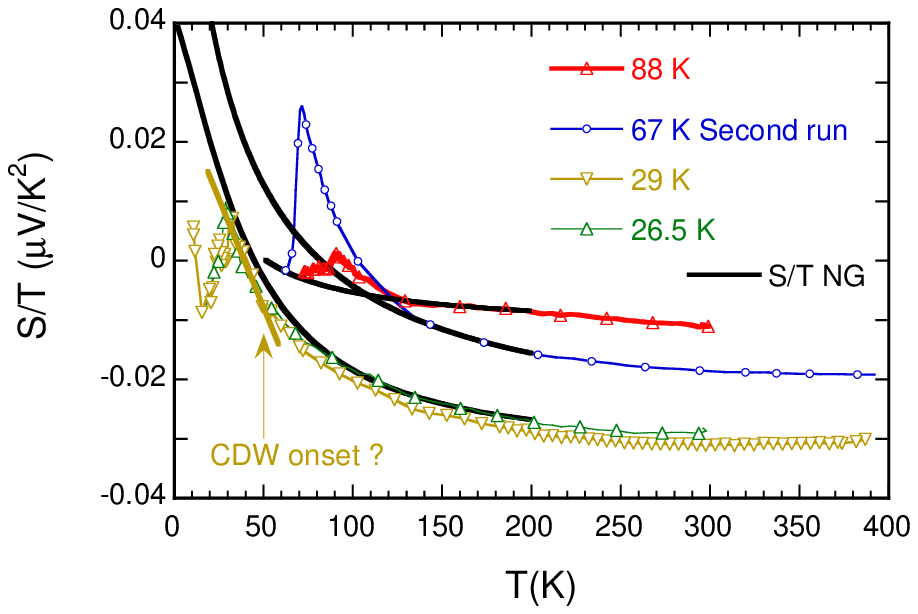}
\caption{$S_a(T)/T$ \textit{vs} $T$ for $T_c$= 88~K, 67~K, 29~K and 26.5~K together with fits to Eq.~\ref{NGrule} for $T_c=$88~K, 67~K and 26.5~K, solid lines, with parameters given in Table~\ref{rhotab}.} \label{SoverTandNG}
\end{figure}

For $T_c\leq$ 67~K the parameters $S_0$, $S_1$, and $S_2$, all assumed to be proportional to $T$, were found by fitting the TEP data in Fig.~\ref{TEPall}(a) to Eq.~\ref{NGrule}. The range of fit was very limited, to below 200~K because of deviations from Eq.~\ref{rhopoly} at higher $T$ and for the two higher $T_c$ values to above 150~K, because of the onset of the CDW. Partly as a result of this, only the parameter $S_2$ representing the contribution to the TEP from the $T^2$ resistivity term is well-defined, although the term $S_0$ from impurity scattering must be positive. This procedure did not work for $T_c= $90~K where it gave the very small value of $\alpha_2$ shown in Table~\ref{rhotab} and a negative value for $\alpha_0$. For this crystal we therefore set $\alpha_0 = 6.4\pm 0.4 \mu \Omega$-cm, the average for the other three $T_c$ values, giving the values of $S_1$ and $S_0$ shown in Table~\ref{rhotab}. For $T_c$ = 88~K and 67~K $S_0$ must be at least 0.1$\mu$V/K$^2$ which is unexpectedly large. In Fig.~\ref{SoverTandNG} it can seen that the calculated $S_{NG}$ curves for $T_c$ = 67~K and 26.5~K, agree well with the data giving $S_2=-0.033T\mu$V/K, of the same sign and approximately 3 times larger than $S_1$ for $T_c$= 90~K. We can therefore conclude that the energy dependence of the $T^1$ scattering term has the same sign as that of the $T^2$ term, which is generally attributed to electron-electron scattering, but is significantly smaller. Measurements in which the superconductivity is suppressed by a magnetic field and on crystals with different residual resistivity, possibly controlled by doping with in-plane impurities such as Zn, would provide a further test of this picture. The magnitude of $S_2$ is discussed below. Because of momentum conservation the electron-electron scattering contribution to $\rho(T)$ usually depends on Umklapp scattering processes which will generally tend to be stronger for larger energies. This implies that for electrons, the electron-electron scattering term $S_2$ will be positive and for holes it will be negative in agreement with experiment for Tl2201. Within this NG picture the changes in TEP below 160~K in Fig.~\ref{TEPall}(a) as $p$ is increased are caused by the increase in the ratio $\alpha_2T^2/\alpha_1T$ with doping $p$ and the fact that $-S_2$ is larger than $-S_1$.

\subsection{Possible anisotropy around the Fermi surface, scenario B}
In the previous sections it was assumed implicitly that the $T^1$ and $T^2$ scattering rates do not vary appreciably around the Fermi surface, so that their contributions to $\rho(T)$ can be added in the same way as for electron-phonon and electron-impurity scattering. This is only approximate because there is experimental evidence from angle-dependent magnetoresistance (ADMR) of the $c$-axis resistivity~\cite{French2009} that for crystals with $T_c$ = 20~K, 17~K and 15~K, the anisotropy, (the parameter $\alpha$ in Ref.~\cite{French2009}) increases from 15$\%$ at 20~K to 40$\%$ at 110~K. Recently, new short-range CDW signals have been identified in RIXS studies of many cuprate families~\cite{Ghiringhelli_rev_2021}. These are present for a wide range of $p$, including the ``strange metal'' region, $p>0.19$, that is relevant for Tl2201 although such short range signals have not yet been observed for Tl2201. They are short-ranged in real space, and therefore broader in reciprocal space and are visible to higher $T$ than the CDWs that have reasonably well-defined, often incommensurate, wave vectors $\underline{Q}$. It has been argued that~\cite{CDF_rho2021} that scattering from these short-range charge density fluctuations~(CDF) are responsible for the $T^1$ term in $\rho(T)$ and because the CDF have a wide range of $\underline{Q}$ vectors they would indeed give isotropic scattering. So if CDF could be observed in Tl2201 crystals a good experimental test of this approach would be to compare their intensities with the size of the $T^1$ terms in $\rho(T)$ for different values of $p$.

Calculations by Rice \emph{et al}~\cite{Rice2017} are more consistent with scenario B. These are based on the theory of Yang, Rice and Zhang~\cite{YRZ2006} and show that Umklapp scattering between hot spots gives rise to a $T^1$ scattering rate in a very limited region of $\underline{k}$-space and a corresponding $T^1$ term in $\rho(T)$. They are not based on the usual Boltzmann transport equation. Somewhat paradoxically, in an earlier paper using the variational principle~\cite{Ziman1960} to solve the Boltzmann equation, Hlubina and Rice~\cite{Hlubina1995} showed that the $T^2$ term will dominate, although of course at low enough $T$ any non-zero $T^1$ term must become larger than a term going as $T^2$. In this scenario $1/\tau$ at or near the hot spots is likely to be close to the Planck value because for $p\simeq$ 0.19 where any $T^2$ term is very small, as shown for example in Table~\ref{rhotab}, $1/\tau$ is somewhat larger than the Planck value. It may be possible to distinguish between scenarios A and B by measuring the resistivity and London penetration depth on several crystals with similar $T_c$ values. Scenario B provides a more direct link between the $T^1$ term and the superfluid density, emphasised recently in Ref.~\cite{Culo2021}, than scenario A. However unlike scenario A it would not necessarily account for Matthiessen's rule, the Nordheim-Gorter rule or the ADMR results. One possible test would be to see if either scenario can account for the ``vertical scaling'' of the electronic specific heat~\cite{Radcliffe1996} in applied magnetic fields.

\subsection{Comparison with the TEP of other correlated electron materials}
The low $T$ TEP of many correlated electron materials, including some where the $T^2$ term in $\rho(T)$ is dominant, has been shown~\cite{Behnia2004} to obey the following simple empirical formula:

\begin{equation}
\label{Behnia}
\frac{S}{T} = \frac{\gamma q}{N_{AV}e}
\end{equation}
Here $\gamma$ is the Sommerfeld coefficient, $N_{AV}$ Avogadro's number and $e$ the magnitude of the electronic charge. The Faraday number $N_{AV}e = 9.6\times{}10^4$ Coulombs/mole. For the simple case of free electron dispersion and 1 electron per formula unit, $q =-1$ so taking $\gamma$ = 0.6 mJ/gm-at/K$^2$~\cite{Wade1994} gives $S = -13~\mu$V/K at 160~K, the end of the linear regions in Fig.~\ref{resdat}(b). This will be reduced by a factor $1+p\simeq1.3$ by the larger carrier concentration and by a factor 2/3 for a constant carrier mean free path~\cite{Behnia2004} giving $-6.7~\mu$V/K at 160~K. The value of $-S_2 = 0.033T\mu$V/K in Table~\ref{rhotab} gives a TEP of $-5.3~\mu $V/K at 160~K. So the $T^2$ electron-electron scattering term in our TEP data for Tl2201 crystals agrees well with findings~\cite{Behnia2004} for many other correlated electron materials, while the contribution from the $T^1$ term is approximately a factor 3 smaller.

\section{Summary}
We have argued that the presence of CDWs can be seen in the in-plane TEP of Tl2201 crystals and in earlier data~\cite{OCT1992} for polycrystalline samples where the measured TEP is very similar to $S_a(T)$ for similar values of $p$. The Nordheim-Gorter rule seems to provide an adequate description of the TEP contributions arising from the $T^1$ and $T^2$ terms in the in-plane electrical resistivity. The magnitude of the TEP contribution from the $T^2$ term agrees with that for other correlated electron conductors~\cite{Behnia2004} while the contribution from the $T^1$ term is $\sim{}\!3$ times smaller. If the $T^1$ term were truly anomalous one might perhaps have expected a larger difference. Within a simple cylindrical Fermi surface approximation, taking into account the known value of $m^*$, the $T^1$ terms for the crystals with $T_c$ of 40~K and 26.5~K correspond to scattering rates $1/\tau$ that are significantly smaller than the Planckian value $k_BT/\hbar$, possibly indicating that $\tau$ is anisotropic. A strongly anisotropic scattering rate could provide a direct way of understanding the apparent correlation between the magnitude of the $T^1$ term and the superfluid density emphasised recently~\cite{Culo2021}, namely only those carriers with a $T^1$ scattering rate would be paired. But there could be difficulties in accounting for Matthiessen's rule, the Nordheim-Gorter rule and the ADMR work which suggests a small anisotropy at low $T$~\cite{French2009}. Also an extremely strong variation in $V_{kk'}$ might not be consistent with $d$-wave behaviour. Somewhat unexpectedly, near 160~K, the $T^2$ term corresponds to a scattering rate $1/\tau \gtrsim k_BT/\hbar$ for all values of the hole concentration.

%\section*{Appendices}
\appendix
\setcounter{section}{1}

\subsection{Method used for TEP}
\begin{figure}
\includegraphics[width=80mm,height=60mm]{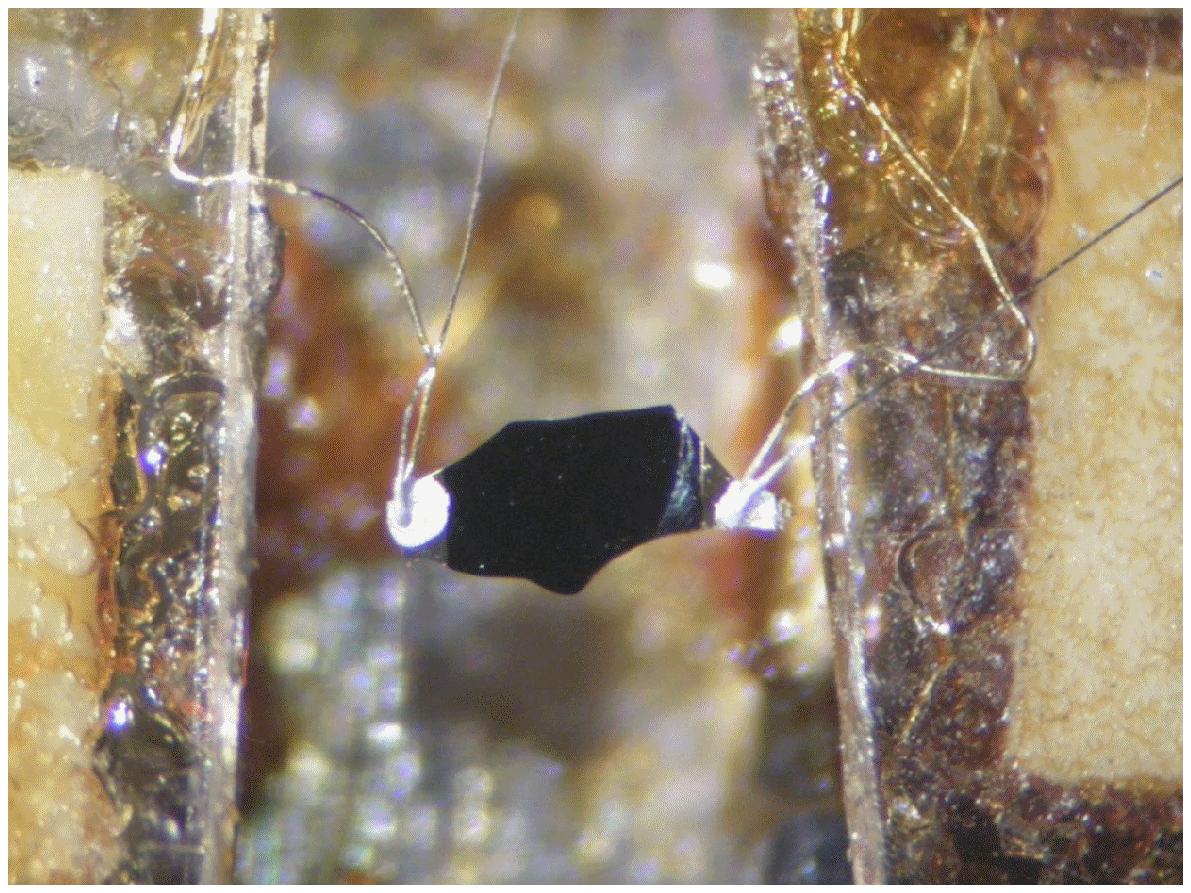}
\label{photo}
\caption{Photo of the 1.5 mm long Tl2201 crystal used for TEP measurements. 50 $\mu$m diameter Au wires and 25 $\mu$m diameter Chromel wires are spot-welded together to form differential thermocouples that are attached to Au pads on the crystal with DuPont 4929N silver paint. The Au wires are heat sunk to sapphire plates with General Electric GE-7031 low temperature varnish.}
\end{figure}

The basic method used for the TEP is similar to that originally used for sintered polycrystalline bars in~\cite{OCT1992}, but because of the small size of the crystals, and the fact that thicker crystals or thin films on a crystalline substrate have high thermal conductance, Au reference wires and Chromel wires forming two differential thermocouples are attached to Au pads on the ends of the crystal as shown in the photograph in Fig.~A1. These wires extend for a few centimetres before being soldered to Cu wires heat sunk to a high thermal conductivity Cu block. A temperature difference~($\Delta T\lesssim1$~K) is produced by alternately heating two sapphire blocks to which the thermocouple wires are attached with General Electric GE-7031 low temperature varnish and the Seebeck coefficient of the sample $S_{\textrm{sample}}$ is given by eliminating $\Delta T$ from the equations:
\begin{equation}
\label{TEP_Eq1}
 {V_{\textrm{Au-Au}} = (S_{\textrm{sample}}-S_{\textrm{Au}})\Delta T}
 \end{equation}
\begin{equation}
\label{TEP_Eq2}
{V_{\textrm{Ch-Ch}} = (S_{\textrm{sample}}-S_{\textrm{Ch}})\Delta T}
\end{equation}
 Here the voltages across the Au wires, $V_{\textrm{Au-Au}}$ and the Chromel wires, $V_{\textrm{Ch-Ch}}$ are measured using a Keithley Instruments digital nanovoltmeter and a home-built nanovolt amplifier respectively, and their ratio obtained by making a least squares linear fit to a plot of $V_{\textrm{Ch-Ch}}$ \textit{vs} $V_{\textrm{Au-Au}}$ for $\pm\Delta T$ and two points with $\Delta T=0$. $S_{\textrm{Ch}}$ is the Seebeck coefficient of Chromel taken from published standards and shown in Fig.~A2. The same reel of $50~\mu$m diameter Au wire was always used and $S_{\textrm{Au}}$, also shown in Fig.~A2 was obtained by measuring a strip of pure Pb which has a low TEP~\cite{PbTEP}.

\begin{figure}
\includegraphics[width=80mm,height=60mm]{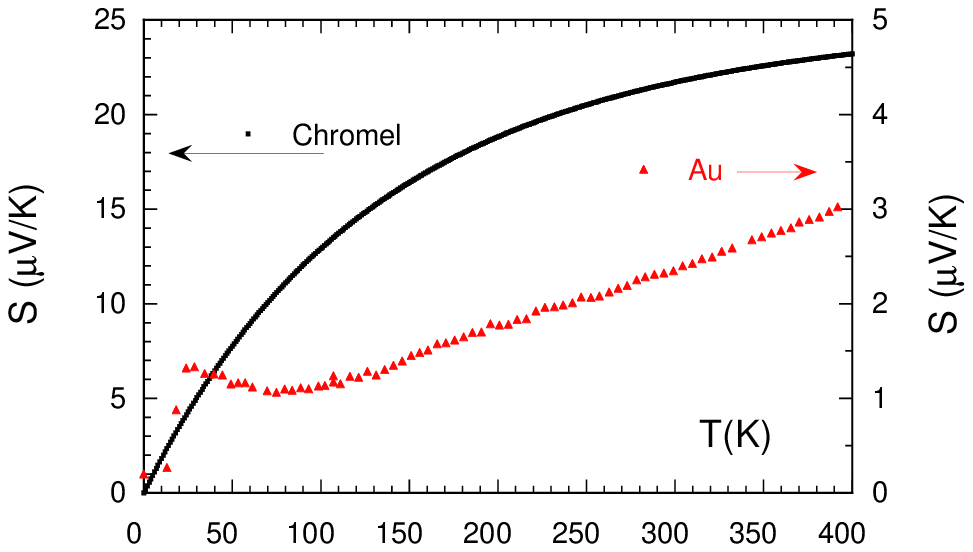}
\label{AuCh}
\caption{Thermopower of Chromel and Au wires used in the present work.}
\end{figure}
\begin{figure}
\includegraphics[width=80mm,height=90mm]{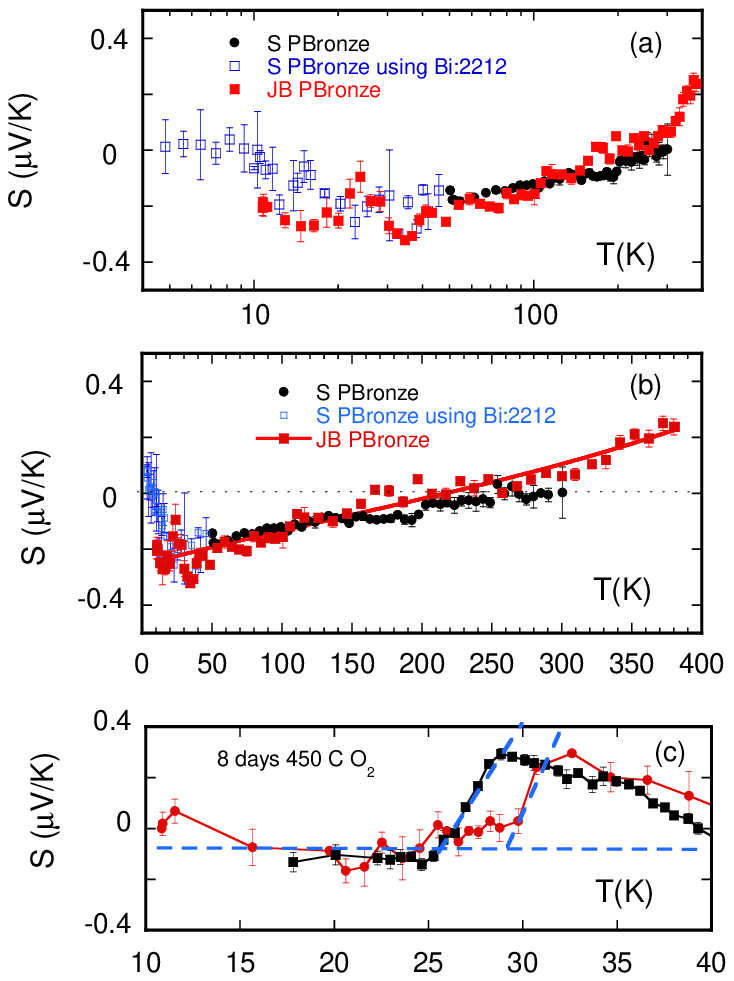}
\label{PBronze}
\caption{(a) and (b) TEP of Lake Shore Cryotronics phosphor bronze alloy wire on logarithmic and linear temperature scales respectively. The most recent data labelled JB PBronze is compared with earlier measurements, including one data set where the superconducting Bi2212 sample had zero TEP below 85~K. (c) Low $T$ data for the Tl2201 crystal after making a small correction of $\sim{}\!0.1~\mu$V/K, given by the data in~(a). The dashed lines show how $T_c$ values of 26~K and 29~K are determined. }
\end{figure}
 In the past further checks have been made by measuring a superconducting cuprate, Bi2212 with $T_c\simeq$~90~K, and a piece of 127~$\mu$m diameter phosphor bronze alloy wire purchased from Lake Shore Cryotronics which has a conveniently low TEP. In the present work a piece of the same wire was measured again and compared with earlier data, as shown in Figs.~A3(a) and (b). Small changes in the TEP of Au can be induced by bending (cold working) the wire which will change the phonon drag component and the electron diffusion component from any Fe impurities. Measuring the phosphor bronze wire is a convenient method of correcting for such changes. Fig.~A3(c) shows the TEP data from Fig.~\ref{TEPall}(a) after making a small correction of $\sim{}0.1~\mu$V/K corresponding to the difference in the TEP of the phosphor bronze wire at low $T$ in earlier and more recent measurements shown in Fig.~A3(a). It can be seen that $T_c$ rises from $26 \pm{}1$ to $29 \pm{}1$~K after one month at room temperature.

\textbf{Acknowledgments}This work was funded by the EPSRC (U.K.) grant number EP/K016709/1. We are grateful to Prof.~N.~E.~Hussey for helpful comments on the first draft of the manuscript.

\end{document}

\title[Supplementary Information: Thermoelectric power of overdoped Tl2201 crystals]{Supplementary Information: Thermoelectric power of overdoped Tl2201 crystals: Charge density waves and $T^1$ and $T^2$ resistivities}

\author{ J R Cooper$^1$, J C Baglo$^2$, C Putzke$^3$\footnote{Present address: Max-Planck-Institute for the Structure and Dynamics of Matter,
Luruper Chaussee 149, Geb. 99 (CFEL), 22761 Hamburg,Germany}, A Carrington$^4$}

\address{$^1$ Cavendish Laboratory, University of Cambridge, J. J. Thomson Avenue, Cambridge, CB3 0HE, United Kingdom}
\address{$^2$ D\'{e}partement de physique, Institut quantique, Universit\'{e} de Sherbrooke, Sherbrooke, Qu\'{e}bec, Canada}
\address{$^3$ Institute of Materials, \'{E}cole Polytechnique F\'{e}d\'{e}ral de Lausanne, Lausanne, Switzerland}
\address{$^4$ H.\ H.\ Wills Physics Laboratory, University of Bristol, Bristol, BS8 1TL, United Kingdom}
\ead{jrc19@cam.ac.uk, Jordan.Baglo@USherbrooke.ca}

\section{Results of ACS and annealing studies} 
\begin{figure}
\includegraphics[width=80mm,height=100mm]{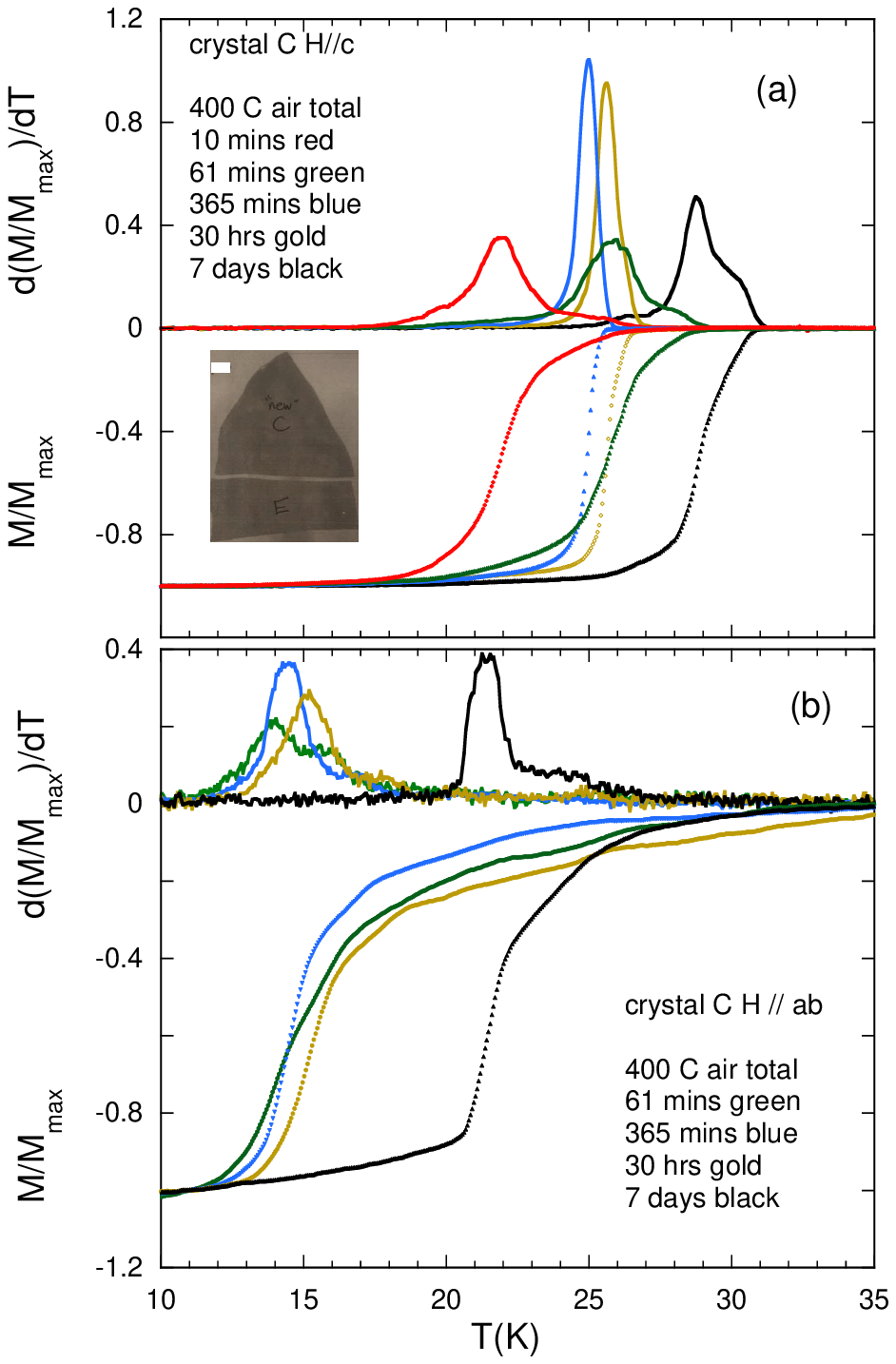}
\caption{Normalized ACS signals $M/M_\mathrm{max}$ and their derivatives $d(M/M_\mathrm{max})/dT$ \textit{vs} temperature for crystal C after various total annealing times in flowing air at 400\,$^{\circ}$C. (a) for $H\|c$ and (b) for $H\|ab$. An image of crystal C, which broke into two pieces, C$'$ and E$'$ after the 365 minute anneal is also shown. The length of the rectangular piece, E$'$ is 1.1\,mm.}
 \label{crystalC}
 \end{figure}

\begin{figure}
 \includegraphics[width=80mm,height=70mm]{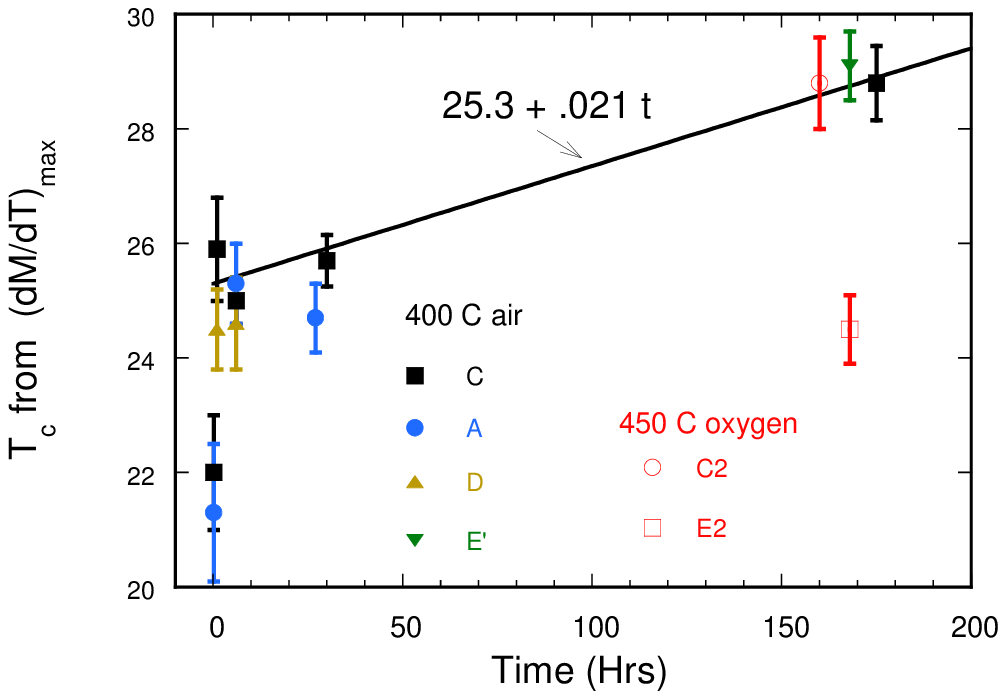}
%\end{verbatim}\normalsize
\caption{$T_c$ values obtained from peaks in derivative plots, $d(M/M_\mathrm{max})/dT$ for $H\|c$ for all crystals measured in this work \textit{vs} total annealing time in air at 400\,$^{\circ}$C or 5N oxygen at 450\,$^{\circ}$C. The total lengths of the error bars correspond to the full widths at the half maxima of these peaks. The peaks are particularly narrow after 6 hours and 30 hours as also seen in Figure~\ref{crystalC}.}
\label{TcTime}
\end{figure}

In this document we refer to the in-plane direction as the $ab$ plane or the $a$ direction interchangeably even though all or our crystals probably have significant excess Cu ions arising from Tl loss during growth and are therefore tetragonal. Normalized ACS data is shown in Figure~\ref{crystalC} for the same crystal C after various annealing times at 400~$^{\circ}$C in flowing air followed by rapid cooling. The data shown and the derivatives were obtained by making sliding averages (smoothing) of the raw data over 20 points, i.e.\ $\pm$0.6\,K. Values of $M_\mathrm{max}$ are given in Table~\ref{crystals}. The out-of-phase signal was also recorded; after making a correction for a small in-phase component we could see that it is very similar to the derivatives of the in-phase component. For $H\|c$ the peaks in the derivatives are used to obtain the $T_c$ values shown for this crystal and all others measured in this work in Figure~\ref{TcTime}. In contrast, and as discussed in the following section, the peaks for $H\|ab$ are broader and at lower temperatures although there is still a significant diamagnetic response up to higher $T$.
\begin{figure}
\includegraphics[width=80mm,height=100mm]{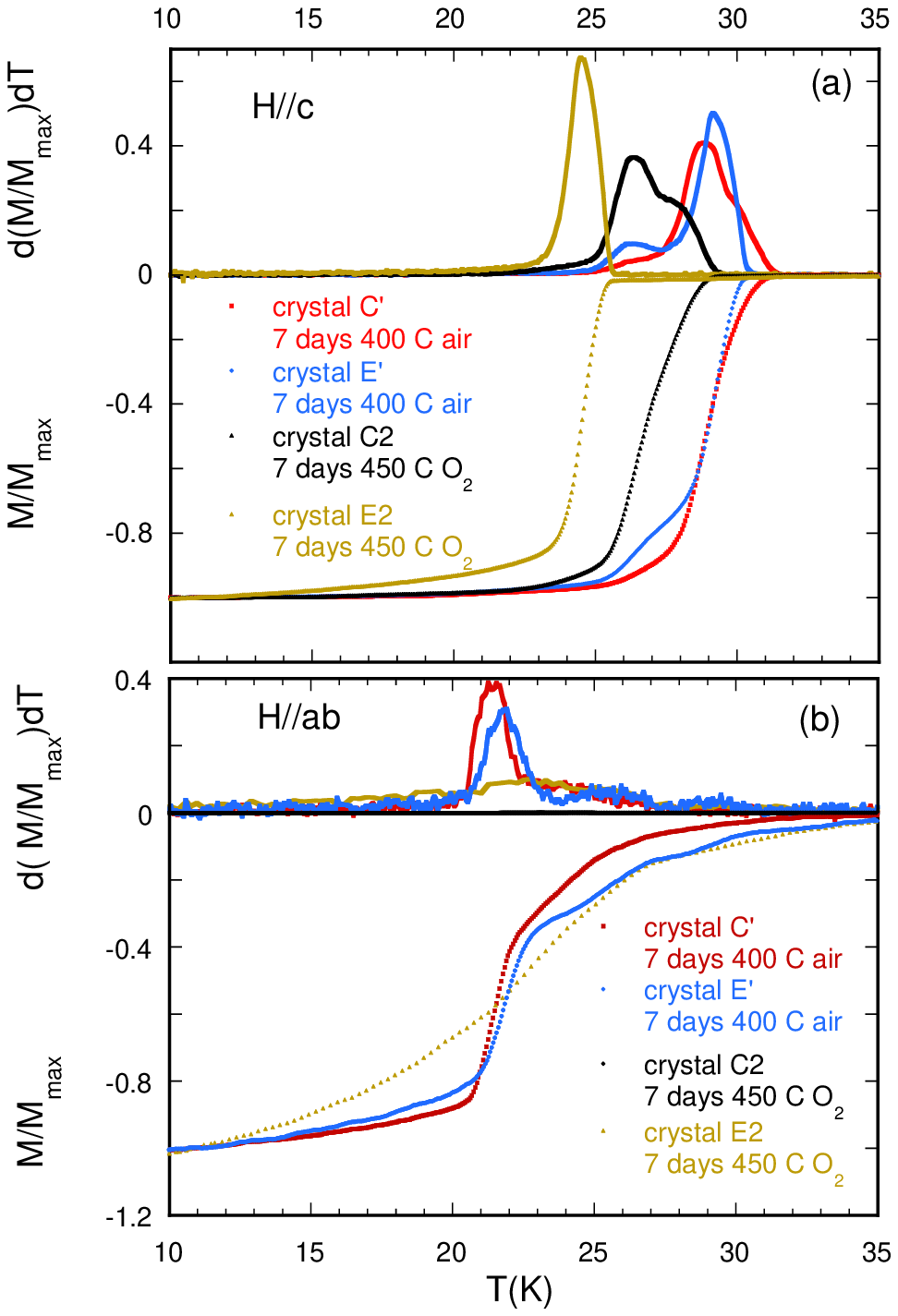}
\caption{Normalized ACS signals $M/M_\mathrm{max}$ and their derivatives $d(M/M_\mathrm{max})/dT$ \textit{vs} temperature for all crystals that were annealed for 7 days. Crystals C$'$ (part of C), E$'$ (part of C) and C2 are from the same preparation batch~(Oct.\ 2014) while E2 is from a later batch~(Nov.\ 2016). C$'$ and E$'$ have different areas; see Table~\ref{crystals}.}
\label{all7days}
\end{figure}
\begin{figure}
\includegraphics[width=80mm,height=80mm]{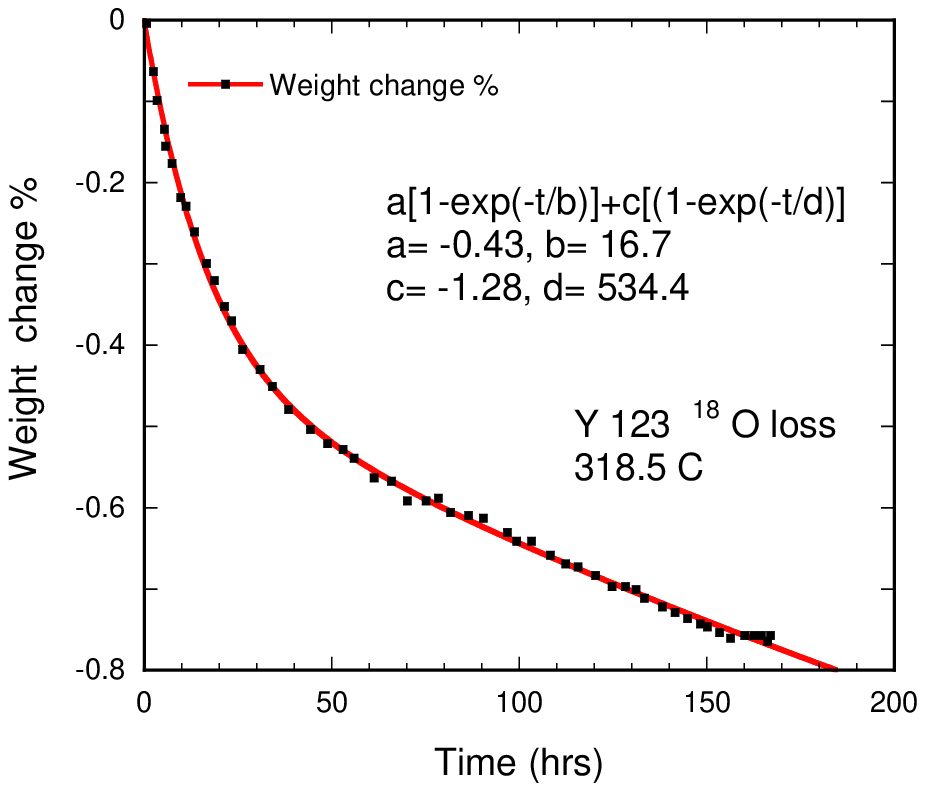}
\caption{Thermogravimetric data from ~\cite{Conder2001} for back-exchange of natural oxygen into YBa$_2$Cu$_3$O$_7$ that was originally fully loaded with $^{18}$O, plus a fit involving two diffusion processes with different time constants.} \label{Conderfig}
\end{figure}
\begin{figure}
\includegraphics[width=80mm,height=100mm]{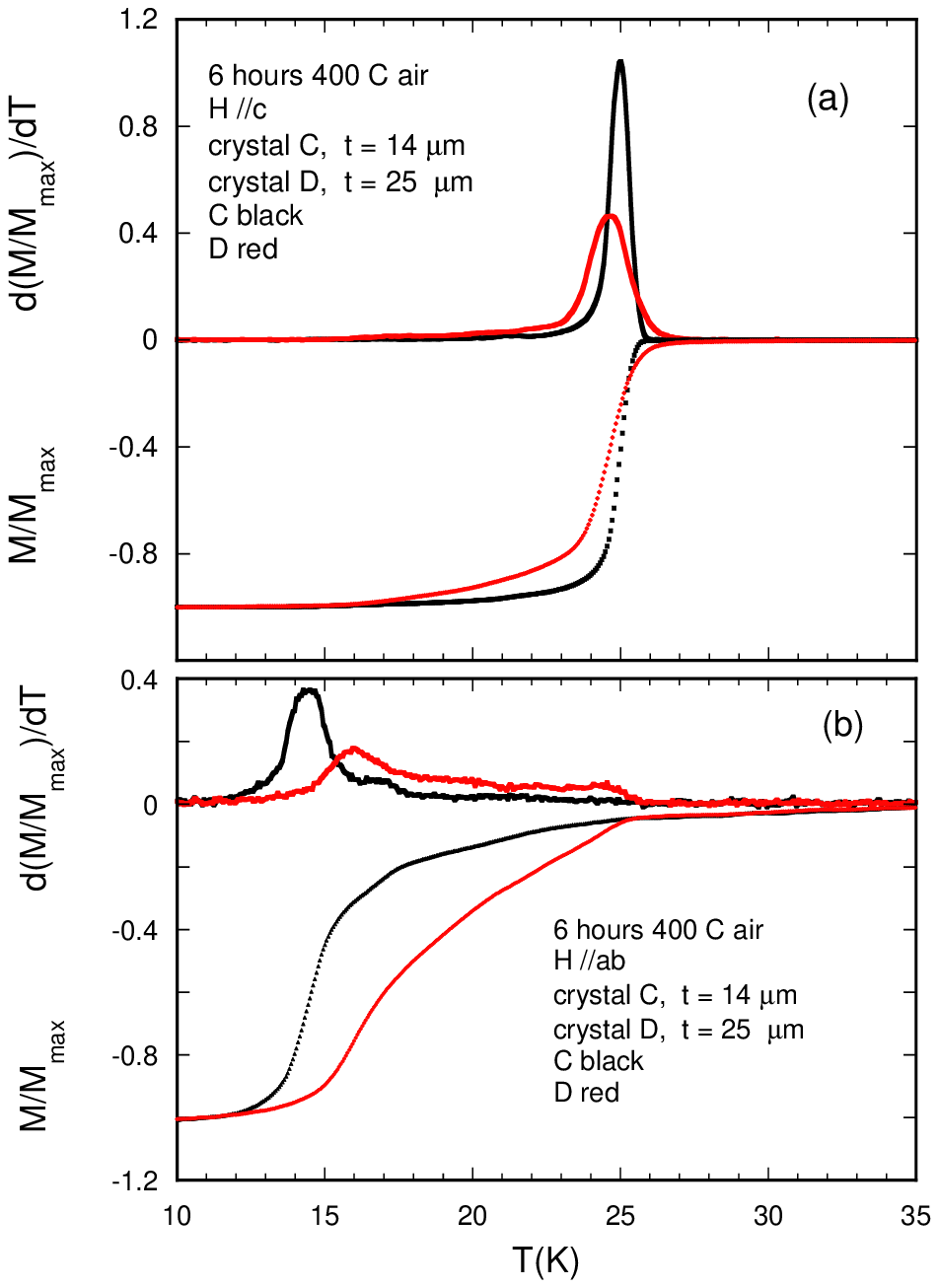}
\caption{Normalized ACS signals $M/M_\mathrm{max}$ and their derivatives $d(M/M_\mathrm{max})/dT$ \textit{vs} temperature for two crystals C and D from the same preparation batch but of different thicknesses that were annealed for 6 hours in flowing air at 400\,$^{\circ}$C. These data suggest that in-plane diffusion is dominant (see text).}
\label{FigC_D}
\end{figure}
\section{Discussion}
Any excess oxygen is believed to be accommodated at interstitial sites in Tl2201~\cite{Aranda1995}, as is probably also the case for Bi2212~\cite{Routbort1994}, but there have not been any studies of diffusion in Tl2201. Possible complicating factors are: (i) $T_c$ of overdoped samples gradually increases with time at room temperature~\cite{Mackenzie1996}, probably because of oxygen clustering or ordering (we saw this effect in our TEP crystal); (ii) tetragonal Tl2201 crystals usually have Cu on the Tl sites, i.e.\ their composition is Tl$_{2-y}$Ba$_2$Cu$_{1+y}$O$_{6+x}$ with $0.05<y<0.15$, and the diffusion of oxygen is affected by alloying~\cite{Tallondiff}; and (iii) the proximity of other crystallographic phases with orthorhombic or monoclinic symmetry~\cite{Attfieldmonoclinic1998}. The diffusion of oxygen in Tl2201 crystals must be reasonably fast because when grown from a flux with typical cooling rates of 60~$^{\circ}$C per hour they are invariably strongly overdoped, with $T_c\simeq{}10$~K or less. Consistent with this, in-plane resistivity measurements on crystals from the same preparation batch as ours~\cite{Putzke2021} showed that for temperatures of 450~$^{\circ}$C and above, annealing times between 30 minutes and 4 hours gave good results.

Returning now to the data for crystal C in Figure~\ref{crystalC}(a) we see that for $H\|c$ the transitions are sharp for anneal times of 6 and 30 hours at 400~$^{\circ}$C. According to the calculations in~\cite{Benseman2008} for one-dimensional diffusion into a long slab of width $b$, the time~$t$ required for homogeneity is given by $t=b^2/D$, where $D$ is the chemical diffusion coefficient. Taking $t$\,=\,10~hours and $b$\,=\,0.7\,mm gives $D$\,=\,1.4\,$\times$\,$10^{-11}$\,m$^2$/s, a factor of 12 more than for Bi2212 in flowing oxygen at 400\,$^{\circ}$C~\cite{Benseman2008}. This supports the arguments in the previous paragraph that much shorter annealing times are needed for Tl2201 than for Bi2212. However as shown in Figure~\ref{crystalC}(a), with longer annealing times $T_c$ continues to increase, approximately linearly with $t$ as shown in Figure~\ref{TcTime}, and after 7 days the derivative shows a broader transition. We believe that this could arise because there is a another chemical diffusion process with a higher activation energy. An example of such behaviour for back-diffusion~\cite{Conder2001} of $^{18}$O out of YBa$_2$Cu$_3$O$_7$ is shown in Figure~\ref{Conderfig}, for which diffusion is faster for O atoms in the Cu-O chains than for other O atoms in the structure. Magnetic susceptibility studies of Tl2201~\cite{Kubo_chi_1991,Wade1995} show that the Curie term attributed to Cu$^{2+}$ ions on the Tl sites increases with oxygen content. It is very likely that oxygen ions are bound to these Cu$^{2+}$ ions~\cite{Peets2010} and their thermally activated release provides another diffusion channel. This would be similar to what happens in YBa$_2$Cu$_3$O$_7$, but an important difference is the small number of Cu ions on Tl sites, that varies from one batch of crystals to another.

A puzzling feature of the data in Figure~\ref{crystalC}(b) is that the peaks in derivatives for $H\|ab$ are up to 9\,K lower than those for $H\|c$ although the onset of superconductivity can be detected in the data for $H\|ab$ at $T_c$ values given by the peaks for $H\|c$. During the experiments we thought that this was a sign of an inhomogeneous oxygen distribution and therefore continued annealing for longer times. As implied above, for $H\|c$ the diamagnetic screening currents will flow in a sheath of depth $\lambda_{ab}(T)$ around the perimeter of the platelet crystal, while for $H\|ab$ these screening currents flow in both the weakly conducting $c$~direction and in the $ab$~plane. One likely interpretation is to ascribe this effect to the presence of macroscopic defects in the layer structure in the $c$~direction, so that the screening currents flowing in the $c$~direction will be attenuated until the superconducting order parameter reaches a certain value. This is supported by the fact that for crystal E2 from a second growth batch, for which there was much less Tl loss during synthesis, there is no distinct peak in $d(M/M_\mathrm{max})/dT$ for $H\|ab$ as shown in Figure~\ref{all7days}. Also the difference in peak temperatures for $H\|ab$ and $H\|c$ for the other crystals, which were all from batch 1, is smaller after annealing for 7 days in either air or oxygen, possibly suggesting that longer anneal times can repair such macroscopic defects.

 Finally, Figure~\ref{FigC_D} shows that for a thicker crystal, D, oxygen diffusion is faster than that out of a thinner one, C. This is of interest because out-of-plane diffusion in other cuprates, e.g.\ YBCO, is at least 100 times slower than in-plane diffusion~\cite{Conder2001}. In the present case this anisotropy must be even larger because the width/thickness ratio of our platelet crystals is 18 (D) or 56 (C) (see Table~\ref{crystals}). So the anisotropy in the chemical diffusion coefficient must be greater than 56$^2$ if diffusion out of crystal C is slower than diffusion out of crystal D. It is also noteworthy that the peak in $d(M/M_\mathrm{max})/dT$ of crystal D for $H\|ab$ is much less marked, supporting our interpretation that it is not an intrinsic property of these crystals.

 \section{Summary}
 The main conclusions from this section are: (i) Anneal times of 10 or so hours at 400 or 450\,$^{\circ}$C are sufficient in order to obtain a uniform hole concentration in crystals that are typically $w$=0.5\,mm wide, confirming the results in~\cite{Putzke2021}. This time will probably increase as $w^2$. (ii) However $T_c$ continues to evolve with longer times. We believe there is an additional diffusion process. This slower process could be linked to the presence of Cu ions on Tl sites, but because the value of $D$ for short annealing times is so much larger than that for Bi2212 crystals, we cannot be sure which mechanism is faster. (iii) ACS measurements with $H\|ab$ are not as informative as we hoped, probably because the crystals studied in detail do have a small number of macroscopic defects in the $c$ direction which are sufficient to impede supercurrent flow between the CuO$_2$ layers without affecting the in-plane electrical conductivity. (iv) Diffusion in the $ab$ planes is much faster than in the $c$ direction.

 Finally we note that the annealing conditions used for the TEP crystal in the main paper were 8 days at 450~C in flowing oxygen ($T_c$ = 26.5~K rising to 29~K after one month at 297~K and then an hour or so at 390~K during measurement), 6 days in flowing argon containing 2.9 ppm O$_2$ at 450~C ($T_c$ = 67~K) and 6 hours at 500~C and 6 x 10$^{-6}$ bars ($T_c$ = 88~K).

\begin{table}
\caption{\label{crystals}Parameters of crystals measured. Rect.= rectangular, tri.= triangular shape. Areas $A$ determined using AxioVision software, except for E2 and C2 where they were estimated from photographs. Thicknesses $t$ in parentheses estimated by requiring $M_\mathrm{max}^{H\|ab}$\,=\,0.23\,$\Omega$/mm$^3$; $\pm$ refers to standard deviation for measurements on same crystal and various annealing times. It is found that $M_\mathrm{max}^{H\|c}/A^{3/2} \approx$\,0.066\,$\Omega$/mm$^3$. For a thin disk of radius $a$, one expects $M_\mathrm{max}^{H\|c} \propto a^3$. As explained in the text the units of $M_\mathrm{max}^{H\|ab}$ are $\Omega$/mm$^3$ while those of $M_\mathrm{max}^{H\|c}$ are $\Omega$.}
\begin{tabular}{@{}llllllll}
\br
Crystal&$A$ (mm$^2$)&$l$ (mm)&$t$ ($\mu m$)&$w_\mathrm{avg.}$ (mm)&$M_\mathrm{max}^{\|{}ab}$&$M_\mathrm{max}^{\|{}c}$&Comment\\
\mr
A rect.      & 0.92  & 1.40 & 13 $\pm$ 1 & 0.66  & 0.129             & 68       &\\
A$'$ rect.     & 0.375 & 1.00 & 13 $\pm$ 1 & 0.375 & 0.14              & $^*$16.4 & $^*$13 after 24 h \\
C tri. + rect. & 1.19  & 1.50 & 14 $\pm$ 1 & 0.79  & 0.240 $\pm$ 0.004 & 80.7     &\\
C$'$ tri.     & 0.71  & 1.04 & 14 $\pm$ 1 & 0.68  & 0.224 $\pm$ 0.007 & 37.4     &part of C\\
D tri.      & 0.596 & 1.29 & 25 $\pm$ 5 & 0.46  & 0.248 $\pm$ 0.012 & 30.4     &\\
E$'$ rect.    & 0.494 & 1.1  & 14 $\pm$ 1 & 0.45  & 0.234             & 20.7     &part of C\\
E2 rect.    & 0.256 & 0.64 & (11)       & 0.40  & (0.23)            & 8.3      &\\
C2 tri.     & 0.376 & 0.87 & (22)       & 0.43  & (0.23)            & 23.3     &\\

\br
\end{tabular}
%\end{indented}
\end{table}

\section{Methods used for ACS and annealing studies}

Many groups use low field AC susceptibility~(ACS) or SQUID magnetization measurements to verify the quality of their single crystals. If the measuring field $H$ is along the $c$ direction of a thin platelet crystal then because of the large demagnetizing factor the signal will be larger and easier to measure. For chemically inhomogeneous crystals there is a danger that it will be dominated by a relatively small number of superconducting layers and give sharp changes at an apparent $T_c$ which is not representative of the whole crystal. Oxygen diffusion is usually much faster in the CuO$_2$ planes~\cite{Routbort1994}. For slightly OD crystals non-uniform in-diffusion of oxygen will show up in systematic measurements with $H\|ab$ and can be used to determine the appropriate chemical diffusion coefficient~\cite{Benseman2008}. For these reasons we have measured the ACS of our platelet crystals, all of whose dimensions are given in Table~\ref{crystals}, for both $H\|c$ and $H\|ab$. Initially the ACS measurements were regarded as only giving qualitative information, so we used a small homemade coil set which was attached to the sample probe of an Oxford Instruments flow cryostat generally used for zero field resistivity measurements. The coil filling factor is much better than more standard systems where the coils are held in a liquid helium bath, e.g.~\cite{Pinteric2000}. Assuming that the noise level does not depend on the size of the crystal then our method give $\sim$10 times less noise, but the empty coil signal is more $T$-dependent. Somewhat surprisingly the sensitivity for a measuring field of 0.1\,G\,rms is better than 10$^{-9}$\,emu\,rms.

Our coil set was made up of a primary coil wound in clear Stycast 1266 epoxy resin on a 4.18\,mm external diameter quartz tube that was specially made for electron spin resonance work and had no detectable magnetic impurities. Two balanced secondary coils were wound on the outside of the primary as done in some early work~\cite{Cooper1990} on Bi2212 crystals. This method takes advantage of the very low thermal expansion of fused quartz. Here we used a measuring frequency of 935.59\,Hz and a primary current of 0.9862\,mA\,rms, which gave an AC field of 0.11\,G rms, and lock-in detection via a 50:1 Princeton Applied Research low-noise transformer. The signal with no sample in the coils was measured separately and subtracted from the signal with a sample. It varied with $T$ because the sample platform for resistivity measurements is made of high purity copper which gave some eddy current losses. Initially there were additional sporadic jumps in this signal because of a small amount of paramagnetic oxygen leaking into the sample space. All data shown here were obtained on warming from approximately 4\,K at 0.5\,K/min. A rectangular Pb$_{0.6}$Sn$_{0.4}$ superconducting strip of dimensions 2.56 $\times$ 1.05 $\times$ 0.197\,mm$^3$ gave a diamagnetic signal of 0.197\,$\Omega$/mm$^3$ with $H$ parallel to the long side. We use these units because the measurement software was the same as that used for electrical resistance, although of course there was a 90\,$^{\circ}$ phase shift. Unfortunately we do not know the residual resistivity of this alloy. It must be two-phase with Pb- and Sn-rich regions~\cite{soft_solder_1979}, and a similar alloy Pb$_{0.3}$Sn$_{0.7}$ has a residual resistivity $\sim{}0.4\,\mu\Omega$-cm, giving a skin depth of $\sim$1\,mm. So it is very likely that the discrepancy of just over $-10\%$ with the data for Tl2201 crystals in Table~\ref{crystals}, typically 0.23\,$\Omega$/mm$^3$ for $H\|ab$, is caused by this effect.

The crystals were attached using Apiezon N grease to 3 mm diameter quartz rods that fitted snugly into the primary coil, one with a flat end for $H\|c$ and the other, with a flat on the side of the rod, for $H\|ab$. The grease was removed by washing in trichlorethylene and then heptane before the next annealing step. Annealing was carried out in a mass spectrometer-tested leak-tight horizontal quartz tube furnace with flowing 5N purity standard air or 5N oxygen gas. The exit of the tube was connected to a bubbler via an oxygen meter with a limiting resolution of 0.1\,ppm. The crystals were held in a small gold boat attached to a Chromel wire that could be pulled out of the furnace without opening the tube, using an small internal iron plate and an external permanent magnet. They were then cooled rapidly by pouring liquid nitrogen on tissue paper wrapped around the tube. In most cases photographs of the crystals were taken and their areas determined using Zeiss AxioVision image processing software; these areas and crystal thicknesses are summarized in Table~\ref{crystals}.

In the work on Bi2212~\cite{Cooper1990} a crystal was cut into two and then four pieces so that measurements with $H\|ab$ could be used to measure the London penetration depth~($\lambda_c$) for screening currents flowing in the weakly conducting $c$ direction. For Tl2201 the situation is less simple because the conductivity~($\sigma$) anisotropy is smaller, with $\sigma_{aa}/\sigma_c\simeq 1000$ at room temperature~\cite{Hussey1996}, so from microwave measurements~\cite{Broun2013} which give $\lambda_{aa}(T=0)$\,=\,0.26\,$\mu$m for a Tl2201 crystal with $T_c$\,=\,25\,K we find $\lambda_c(T=0)=8.2~\mu$m. For a rectangular crystal of width $w$ and thickness $t$ at low $T$ we expect the diamagnetic signal with $H\|ab$ to be reduced by a factor $(1-2\lambda_c(0)/w)(1-2\lambda_{ab}(0)/t)$. So for typical dimensions $w$\,=\,500\,$\mu$m and $t$\,=\,14\,$\mu$m the two factors in brackets are 0.97 and 0.96 respectively. Similar considerations involving a narrow ``normal'' sheath of depth $\lambda_c$ or $\lambda_{aa}$ around the circumference of a non-rectangular crystal apply at low $T$ for $H\|ab$, but when the screening diamagnetism is small a more precise analysis, probably using commercially available software for finite element analysis, would be needed.